\newif\ifjournal

\ifjournal
\documentclass[final,3p,times]{elsarticle}
\else
\documentclass[final,5p,times,twocolumn]{elsarticle}
\fi

\usepackage[table]{xcolor}% http://ctan.org/pkg/xcolor
\usepackage{amsmath,amsthm}
\usepackage{amssymb}
\usepackage{amscd}
\usepackage{textcomp}
\usepackage{hyperref}
\usepackage{fancyvrb}
\usepackage{wasysym}

\usepackage{booktabs}
\usepackage{lineno}

\graphicspath{{graphics/}}
\usepackage{listings}

\newcommand{\fastest}{\cellcolor{green!25}}
\newcommand{\slowest}{\cellcolor{red!25}}

\usepackage{color}
\definecolor{red}{rgb}{1,0,0}
\definecolor{blue}{rgb}{0,0,1}
\definecolor{green}{rgb}{0,0.5,0}
\definecolor{magenta}{rgb}{1,0,1}

\newsavebox{\ieeealgbox}

\newcommand{\be}{\begin{equation}}
\newcommand{\ee}{\end{equation}}
\newcommand{\bea}{\begin{eqnarray}}
\newcommand{\eea}{\end{eqnarray}}
\newcommand{\bal}{\begin{align}}
\newcommand{\eal}{\end{align}}
\newcommand{\nn}{\nonumber}

\journal{Elsevier}

\begin{document}

\ifjournal
\linenumbers
\fi
\begin{frontmatter}

\title{Linear Optimal Power Flow Using Cycle Flows}

\author[fias]{Jonas H\"orsch}
\author[penn]{Henrik~Ronellenfitsch}
\author[fzj,koln]{Dirk~Witthaut}
\author[fias]{Tom Brown\corref{cor1}}
\ead{brown@fias.uni-frankfurt.de}

\cortext[cor1]{Corresponding author}

\address[fias]{Frankfurt Institute for Advanced Studies, 60438 Frankfurt am Main, Germany}
\address[penn]{Department of Physics and Astronomy, University of Pennsylvania, Philadelphia PA, USA}
\address[fzj]{Forschungszentrum J\"ulich, Institute for Energy and Climate Research -  Systems Analysis and Technology Evaluation (IEK-STE),  52428 J\"ulich, Germany}
\address[koln]{Institute for Theoretical Physics, University of Cologne, 50937 K\"oln, Germany}

%\date{\today }

\begin{abstract}
Linear optimal power flow (LOPF) algorithms use a linearization of the
alternating current (AC) load flow equations to optimize generator
dispatch in a network subject to the loading constraints of
the network branches. Common algorithms use the voltage angles
at the buses as optimization variables, but alternatives can be
computationally advantageous. In this article we provide a review
of existing methods and describe a new formulation that expresses the
loading constraints directly in terms of the flows themselves, using a
decomposition of the network graph into a spanning tree and closed cycles.
We provide a comprehensive study of the computational performance of
the various formulations, in settings that include
computationally challenging applications such
as multi-period LOPF with storage dispatch and generation capacity expansion.
We show that the new formulation
of the LOPF solves up to 7 times faster than the angle formulation
using a commercial linear programming solver, while another existing cycle-based formulation solves up to 20 times faster,
with an average speed-up of factor 3 for the standard networks
considered here.
If generation capacities are also optimized, the average speed-up rises
to a factor of 12, reaching up to factor 213 in a particular instance.
The speed-up is largest for networks with many buses
and decentral generators throughout the network, which is highly
relevant given the rise of distributed renewable generation and
the computational challenge of operation and planning in such networks.
\end{abstract}


\end{frontmatter}

% --- Content --------------------------------------------------------------------

\section{Introduction}

Optimal Power Flow (OPF) problems can be constructed to find the welfare-maximizing generation and
consumption levels in a network given the physical load flow equations,
branch loading limits and generator cost functions. The full load flow equations are
non-linear and the resulting optimization problem is non-convex, which
makes it both challenging and computationally expensive to find a
global optimum \cite{CAPITANESCU201657}. In transmission networks with sufficient reactive
power compensation, linearizing the load flow equations introduces
only small errors \cite{Purc05,Stot09}, with the benefit that the
Linear OPF (LOPF) can be expressed as a linear problem, whose
convexity guarantees that a local optimum is a global optimum.

LOPF algorithms are principally used in applications with high
computational complexity where it would be impossible to use the full
load flow equations, such as
clearing markets with nodal pricing
\cite{Schweppeetal1988} (particularly with multi-period storage constraints and/or generator unit commitment), determining redispatch measures in markets
with zonal pricing \cite{Burstedde2012}, optimizing dispatch taking
account of contingencies (Security Constrained LOPF (SCLOPF))
\cite{Wood14,CAPITANESCU20111731} and in the long-term optimization of investment in
generation and transmission assets \cite{Latorre,Lumbreras201776}.
Where higher accuracy solutions are required, linear solutions can be
fed as an initial solution into algorithms that use the full non-linear
load flow equations \cite{CAPITANESCU201657}.
LOPF is becoming
more important with the growth of renewable energy, since the
fluctuating feed-in has led to more frequent situations where the
network is highly loaded \cite{Pesc14}. When large networks are
optimized over multiple representative feed-in situations, especially
with discrete constraints on generation dispatch, the LOPF problems
can still take a significant time to solve, despite the linearization of the
problem. Approaches in the literature to reducing the computational
times of LOPF problems include decomposition \cite{736264,1216140,7541033,YAMIN2003101,Wang2017127}, reformulating the problem using Power Transfer Distribution Factors (PTDFs) \cite{HINOJOSA2016391,HINOJOSA2016139} and a parallelizable algorithm using the primal-dual
interior point method \cite{7540826}.

In textbooks \cite{Grai94,Wood14} and major software packages such as
MATPOWER \cite{MATPOWER}, DIgSILENT PowerFactory \cite{PowerFactory},
PowerWorld \cite{PowerWorld} and PSAT \cite{PSAT}, the linearization
of the relations between power flows in the network and power
injection at the buses is expressed indirectly through auxiliary
variables that represent the voltage angles at the buses. In this
paper we introduce a new formulation of the LOPF problem that use the
power flows directly, decomposed using graph theoretic techniques into
flows on a spanning tree and flows around closed cycles in the
network. The new formulation involves both fewer decision variables
and fewer constraints than the angle-based formulation.  We evaluate
the computational performance of the various methods for the LOPF
problem, showing that the cycle-based formulations can solve
significantly faster than the traditional angle-based formulation. We
examine not just the basic LOPF problem, but also applications that
include more computationally challenging multi-period storage
optimization and generation capacity expansion.

Cycle-flow techniques have already been used in \cite{ronellen16} to improve
the calculation times of PTDFs
and to gain a new understanding of the propagation of line outages in
networks \cite{ronellen16b}. The cycle-based LOPF formulation we call the `Kirchhoff formulation' below was used in \cite{Carvalho} for single-period LOPF and in
\cite{Kocuk16}  for single-period LOPF with optimal transmission switching; in contrast
to those papers, here we provide an additional new cycle-based formulation and
benchmark both formulations against established formulations for a different set of computationally-challenging problems: those extending over multiple periods.

In Section \ref{sec:lpf} the different formulations of the linear load
flow are reviewed to prepare for the
introduction of the optimization in Section \ref{sec:lopf}. Extensions
beyond the basic LOPF problem are described in Section
\ref{sec:extensions} and the results of the performance analysis
are presented in Section \ref{sec:results}. Variables are defined in Table \ref{tab:variables}.

\section{Linear load flow formulations}\label{sec:lpf}

\begin{table}[!t]
	\caption{Variable definitions}
	\label{tab:variables}
	\centering
        \small
	\begin{tabular}{@{}lp{5cm}@{}}
\toprule
Variable & Definition \\
\midrule
$i,j \in \{1,\dots N\}$ & Bus labels \\
$s \in \{1,\dots G\}$ & Generation source labels (wind, solar, gas, etc.) \\
$k,\ell \in \{1,\dots L\}$ & Branch labels \\
$c,d \in \{1,\dots L-N+1\}$ & Cycle labels \\
$t \in \{1, \dots T \}$ & Snapshot / time point labels \\
$d_{i,s}$ & Dispatch of generator at bus $i$ with source $s$ \\
$D_{i,s}$ & Available power of generator $i,s$ \\
$l_i$ & Electrical load at bus $i$ \\
$\theta_i$ & Voltage angle at bus $i$ \\
$p_i$ & Total active power injection \\
$\theta_\ell$ & Voltage angle across a branch \\
$f_\ell$ & Branch active power flow \\
$g_\ell$ & Flow on spanning tree (zero if $\ell$ not in tree) \\
$h_c$ & Flow around cycle $c$ \\
$F_\ell$ & Branch active power rating\\
$x_\ell$ & Branch series reactance \\
$K_{i\ell}$ & $N \times L$ incidence matrix \\
$C_{\ell c}$ & $L \times (L-N+1)$ cycle matrix \\
$T_{\ell i}$ & $L \times N$ tree matrix \\
$B_{\ell k}$ & Diagonal $L\times L$ matrix of branch susceptances \\
$\Lambda$ & $N \times N$ weighted Laplacian matrix \\
& $\Lambda = KBK^T$ \\
\bottomrule
	\end{tabular}
\end{table}

%In this section four methods are presented for solving the linear
%load flow. Their computational behaviour is determined by calculations
%that are independent of the $p_i$ that only have to be performed once,
%and solutions of sets of linear equations that have to be performed
%for each new load flow situation determined by the $p_i$. The
%properties of the formulations are summarized in Table
%\ref{tab:loadflow}. The linear equations to be solved are all sparse
%and therefore amenable to fast sparse matrix solution techniques
%and thus solve within milliseconds even for large networks.

The aim of the linear load flow calculation is to calculate the active
power flow $f_\ell$ on each of the branches $\ell=1,\ldots,L$ in terms
of the active power $p_i$ injected or consumed at each of the buses
$i=1,\ldots,N$. In this section four methods are presented for solving
the linear load flow, which lead to different formulations of the LOPF
problem, as discussed in the next section. The different formulations lead to
mathematically identical solutions, as demonstrated in this section.

The linear approximation is valid if all branch resistances $r_\ell$
are negligible compared to the branch reactances $x_\ell$, $r_\ell \ll
|x_\ell|$, reactive power flows may be neglected, all voltage
magnitudes are kept at nominal value and if all voltage angle
differences across branches $\theta_\ell$ are small enough that we can
approximate $\sin\theta_\ell \sim \theta_\ell$. The usefulness of the
linear approximation and the errors thereby introduced are discussed
in \cite{Purc05,Stot09}. If the approximation holds, the real power
over a transmission line $\ell$ is given by
\begin{equation}
  f_\ell = \frac{\theta_\ell}{x_\ell}, \label{eq:ftotsimple}
\end{equation}
where $\theta_\ell$ is the voltage angle difference between the terminal
buses of line $\ell$.

The flows $f_\ell$ are constrained to be physical by the two Kirchhoff circuit laws
for the current and voltage.
Kirchhoff's Current Law (KCL) states that the current injected at each
bus must equal the current withdrawn by the branches
attached to the bus. This law can be expressed using the
incidence matrix $K_{i\ell}$, which has non-zero values $+1$ if branch
$\ell$ starts on bus $i$ and $-1$ if branch $\ell$ ends on bus
$i$.  KCL then reads
\begin{equation}
  p_i = \sum_\ell  K_{i\ell} f_\ell \hspace{1cm} \forall i=1,\ldots, N. \label{eq:kcl}
\end{equation}
KCL directly implies power conservation $\sum_i p_i = 0$ because
$\sum_i K_{i\ell} = 0$ for all lines $\ell$.
KCL provides $N$ linear equations for the $L$ unknown flows $f_\ell$,
of which one is linearly dependent. This is not sufficient to uniquely
determine the flows unless the network is a tree. Hence, $L-N+1$
additional independent equations are needed.

%\begin{table}[!t]
%	\caption{Load flow formulations}
%	\label{tab:loadflow}
%	\centering
%	\begin{tabular}{@{}llr@{}}
%\toprule
%Formulation & One-off & \# Linear equations  \\
%& calculations & to solve per load flow\\
%\midrule
%Angle & $K,\Lambda$ & $N-1$ \\
%PTDF & $K,\Lambda^{*}, PTDF$ & 0 \\
%Kirchhoff &  $K,C$ & $L$ \\
%Cycle & $K,C,T$ & $L-N+1$ \\
%\bottomrule
%	\end{tabular}
%\end{table}

The necessary equations and physicality are provided by the Kirchhoff
Voltage Law (KVL), which states that the sum of potential differences
across branches around all cycles in the network must sum to zero. It
follows from graph theory that there are $L-N+1$ independent cycles
for a connected graph \cite{Dies10}, which provides enough equations
to constrain the $f_\ell$ completely.  The independent cycles
$c\in\{1,\dots L-N+1\}$ are expressed as a directed linear combination
of the branches $\ell$ in the cycle incidence matrix
\begin{equation}
C_{\ell c} = \left\{
   \begin{array}{r l}
      1 & \; \mbox{if edge $\ell$ is element of cycle $c$},  \\
      - 1 & \; \mbox{if reversed edge $\ell$ is element of cycle $c$},  \\
      0     & \; \mbox{otherwise}.
  \end{array} \right.
  \label{eqn:cycle-edge-matrix}
\end{equation}
Then the KVL becomes
\begin{equation}
  \sum_{\ell} C_{\ell c} \theta_{\ell} = 0 \hspace{1cm} \forall c=1,\ldots,L-N+1.  \label{eq:kvl}
\end{equation}
where $\theta_\ell = \theta_i - \theta_j$ is the angle difference
between the two buses $i,j$ which branch $\ell$ connects.
Using equation (\ref{eq:ftotsimple}), KVL can be expressed in terms of the power flows as
\begin{equation}
  \sum_{\ell} C_{\ell c} x_\ell f_\ell = 0 \hspace{1cm} \forall c=1,\ldots,L-N+1. \label{eq:kvlf}
\end{equation}

\subsection{Angle formulation}

Commonly, the linear load flow problem is formulated in terms of the voltage
phase angles $\theta_i, i \in \{1,\ldots,N \}$. Using the incidence matrix the power
flows are expressed as
\begin{equation}
  f_\ell = \frac{1}{x_\ell} \sum_{i} K_{i\ell} \theta_i \hspace{1cm} \forall \ell=1,\ldots,L \label{eq:ftot}
\end{equation}
If the $L\times L$ diagonal matrix $B$ is defined with $B_{\ell \ell} = \frac{1}{x_\ell}$
then the KCL equation (\ref{eq:kcl}) becomes
\begin{align}
  p_i &= \sum_{\ell,k,j} K_{i\ell} B_{\ell k} K_{j k} \theta_j \nn \\
      &= \sum_{j} \Lambda_{ij} \theta_j,
  \hspace{1cm} \forall i=1,\ldots,N, \label{eq:ptot}
\end{align}
using the nodal susceptance matrix matrix $\Lambda = KBK^T$.
In mathematical terms, $\Lambda$ is a weighted network Laplacian \cite{Newm10}.

The Angle formulation thus consists of two consecutive steps to calculate the flows $f_\ell$.
First, equation (\ref{eq:ptot}) is solved to obtain the $N$ voltage angles $\theta_i$. The
equation provides only $N-1$ independent conditions such that we typically fix the
voltage angle at a slack bus as $\theta_0 = 0$. Second, the flows are calculated via
Equation (\ref{eq:ftot}).  KVL is automatically satisfied as all closed cycles are
in the kernel of the incidence matrix such that
\begin{equation}
  \sum_{\ell} K_{i \ell} C_{\ell c}  = 0 \hspace{1cm} \forall c=1,\ldots,L-N+1.
\end{equation}

%\begin{figure*}[tb]
%  \centering
%  \begin{circuitikz}
%  \draw
%  (0,3)
%  to [short,i^=$f_1$,*-*] (3,3)
%  to [short,i>=$f_2$,*-*] (3,0)
%  to [short,i>=$f_3$,*-*] (0,0)
%  to [short,i>=$f_4$,*-*] (0,3);
%  \draw (3,3)   to [short,i^=$f_5$,*-*] (0,0);
%
%  \draw (1.5,-1.2) node{$f_\ell$};
%
%
%  \draw (4,1.5) node{$=$};
%
%  \draw (4,-1.2) node{$=$};
%
%
%  \draw[red]
%  (5,3)
%  to [short,i^=$g_1$,*-*] (8,3)
%  to [short,i>=$g_2$,*-*] (8,0)
%  to [short,i>=$g_3$,*-*] (5,0);
%
%  \draw (6.5,-1.2) node{$g_\ell$};
%
%
%  \draw (9,1.5) node{$+$};
%
%  \draw (9,-1.2) node{$+$};
%
%
%
%  \draw[blue]
%  (10,3)
%  to [short,i>=$$,*-*] (12.8,3)
%  to [short,i>=$$,*-*] (10,0.2)
%  to [short,i>=$h_1$,*-*] (10,3);
%
%  \draw[blue]
%  (13,0)
%  to [short,i>=$$,*-*] (10.2,0)
%  to [short,i>=$$,*-*] (13,2.8)
%  to [short,i>=$h_2$,*-*] (13,0);
%
%
%  \draw (11.5,-1.2) node{$\sum_k C_{\ell,k} h_k$};
%
%\end{circuitikz}
%\caption{
%\label{fig:treecycle}
%Decomposition of flows into a spanning tree part and flows around a basis of closed cycles.
%}
%\end{figure*}

\subsection{PTDF formulation}

For the Power Transfer Distribution Factor (PTDF) formulation
\cite{Wood14} the matrix defining equation \eqref{eq:ptot} is
explicitly inverted to get the angles in terms of the power
injections, and the resulting expression for the angles inserted into
\eqref{eq:ftot} to get a direct linear relation:
\begin{align}
  f_\ell = \sum_i PTDF_{\ell i} \, p_i \hspace{1cm} \forall \ell=1,\ldots,L,
\end{align}
where the PTDF matrix is given by $PTDF = BK^T \Lambda^{*}$. The
pseudo-inverse $\Lambda^{*}$ is used because $\Lambda$ contains a
zero eigenvalue for a connected network. Because KCL is no longer
explicitly enforced, power conservation $\sum_i p_i = 0$ must be added
as an explicit constraint for each connected network.
The need to calculate the explicit pseudo-inverse of $\Lambda$ makes
this slow compared to the Angle formulation for single calculations,
but once the PTDF has been computed, repeated application involves
only matrix multiplication and no equation-solving. However, the PTDF
matrix is typically dense, while $\Lambda$ and $K$ are sparse.

\subsection{Kirchhoff formulation}

In what we call the `Kirchhoff formulation', the linear load flow is
expressed as explicit linear constraints on the flows themselves. To
the $N-1$ independent equations of the KCL equation from \eqref{eq:kcl} we
add the $L-N+1$ constraints of the KVL from \eqref{eq:kvlf}. Together, this provides
a system of $L$ independent equations for the $L$ variables $f_\ell$ and can
therefore be solved.

\subsection{Cycle formulation}
\label{sec:flow-cycle}

In what we call the `Cycle formulation' the flows $f_\ell$ are
decomposed into a superposition of the flows $g_\ell$ on a spanning tree of the network,
which ensure KCL is satisfied, and into cycle flows $h_c$ that flow around
each independent cycle $c$ in the network without altering the power
balance at any bus \cite{ronellen16}. We thus have:
\begin{equation}
  f_{\ell} = g_{\ell} + \sum_c C_{\ell c} h_c . \label{eq:decomp}
\end{equation}

The $g_\ell$ are only non-zero on the $N-1$ edges of a chosen spanning
tree of the connected network.  They are uniquely determined from the
power imbalances by a matrix $T$
\begin{equation}
  g_{\ell} = \sum_{i} T_{\ell i} p_i .
\end{equation}
$T$ is determined by fixing a slack bus and giving $T_{\ell i}$ value
$+1$ if branch $\ell$ is in the directed path in the spanning tree from
$i$ to the slack bus or $-1$ if it is in the directed path but with
reversed orientation \cite{ronellen16}. This guarantees that KCL is satisfied at every
bus given that the power is balanced, $\sum_i p_i = 0$.
Note that $T$ only has to be calculated once for a network and
is independent of the $p_i$. There is freedom both in the choice of spanning tree
and in the choice of the slack bus used to determine the matrix $T$.

The remaining $L-N+1$ degrees of freedom for the cycle flows $h_c$ are
fixed by the $L-N+1$ additional constraints from KVL \eqref{eq:kvlf}
\begin{equation}
\sum_{\ell} C_{\ell c} x_\ell \left(g_{\ell} + \sum_d C_{\ell d} h_d\right) = 0  \hspace{1cm}  \forall c \label{eq:kvlh}
\end{equation}
Solving this equation for the $h_c$ involves solving $L-N+1$ linear
equations. Power networks are not so heavily meshed, typically $L-N+1 <
N-1$, such that this method can be significantly faster than the
Angle formulation \cite{ronellen16,ronellen16b}.

\section{Linear optimal power flow formulations}\label{sec:lopf}

In this section the linear load flow methods from Section
\ref{sec:lpf} are transposed to the linear optimal power flow
(LOPF). In optimal power flow, power plant dispatch is optimized to
minimize dispatch costs, assuming that no branch flows $f_\ell$ exceed
their loading limits $F_\ell$, i.e. $|f_\ell| \leq F_\ell$ \cite{Wood14}.

The factors which control the speed of the solution to the LOPF
problem are now more subtle. They include: i) the number of
optimization variables; ii) the number of
constraints; iii) the sparsity or density of the constraint
matrix; iv) the shape of the feasible space near the optimal point; v) the method used to solve the linear problem. The first three
factors are summarized for each of the formulations in Table
\ref{tab:formulations}.

\begin{table*}[!t]
	\caption{Overview of the different formulations of the LOPF problem
	($N$: number of buses, $L$: number of transmission lines, $G$: number of dispatchable generators)}
	\label{tab:formulations}
	\centering
	\begin{tabular}{@{}lllllll@{}}
\toprule
Formulation & Variables & \# Variables
& \# Equality constraints & \# Inequality constraints & Matrices \\
\midrule
Pure Angle & $d_{i,s}, \theta_i$ & $G+N$ &  $N+1$  &  $G+2L$ & sparse\\
Angle+Flow & $d_{i,s},f_\ell, \theta_i$ & $G+L+N$ & $L+N+1$ &  $G+2L$ &  sparse \\
Pure PTDF & $d_{i,s}$ & $G$  & $1$   &  $G+2L$ &  dense \\
PTDF+Flow & $d_{i,s},f_\ell$ & $G+L$  & $L+1$  &  $G+2L$ &  dense \\
Kirchhoff &  $d_{i,s},f_\ell$ & $G+L$  & $L+1$   & $G+2L$ &  sparse \\
Pure Cycle & $d_{i,s},h_c$ & $G+L - N + 1$  & $L-N + 2 $  & $G+2L$ &  semi-sparse \\
Cycle+Flow & $d_{i,s},h_c,f_\ell$ & $G+2L - N + 1$ & $2L-N + 2 $  & $G+2L$ &  semi-sparse \\
\bottomrule
	\end{tabular}
\end{table*}

The objective function for the LOPF has the generic form
\begin{equation}
   \min_{\{d_{i,s}\}, \{z_a \}} \left[ \, \sum \nolimits_{i,s} c_{i,s} d_{i,s} \right]
   \label{eq:opt-withoutf}
\end{equation}
where $d_{i,s}$ is the dispatch of generator $s$ at bus $i$ and
$c_{i,s}$ is its operating cost. The $z_a$ are auxiliary variables
that implement the network constraints and depend on the problem
formulation (for instance, they would be the voltage angles in the
case of the Angle formulation).

One can also include the line flows $f_\ell$ as explicit optimization
variables. The generic optimization
problem then reads
\begin{equation}
   \min_{\{d_{i,s}\}, \{z_a \}, \{f_\ell\}} \left[ \, \sum \nolimits_{i,s} c_{i,s} d_{i,s} \right]
   \label{eq:opt-withflow}
\end{equation}
All variables and their definitions are listed in Table \ref{tab:variables}.

The optimization must respect several constraints. First, the load $l_i$ at each bus
(which is assumed to be inelastic) must always be met.  The bus power balance is the
difference between generation and the electrical load $l_i$ at the bus
\begin{equation}
  p_i = \sum_{s} d_{is} - l_i \, .
  \label{eq:pi_vs_li}
\end{equation}
If $p_i>0$ then the bus is a net exporter of power; if $p_i < 0$ then the bus is a net
importer of power. Note that $p_i$ is only used to organize the presentation of the
equations and is not an explicit optimization variable.
Second, each generator must dispatch within its available power
\begin{equation}
  0 \leq d_{i,s} \leq D_{i,s} \qquad \forall \; \mbox{generators} \, .
  \label{eq:maxdispatch}
\end{equation}
Third, the real power flows must remain within the loading limits of the lines
\begin{equation}
  |f_\ell| \leq F_\ell \qquad \forall \; \ell = 1,\ldots,L  .\label{eq:thermal}
\end{equation}

It is sometimes desirable to limit the magnitude of the voltage angle
differences $\theta_\ell$ across the branches, to maintain the $\sin
\theta_\ell \sim \theta_\ell$ approximation and avoid voltage
stability problems \cite{kundur1994power}. Since $\theta_\ell = x_\ell f_\ell$, this
constraint has the same form as the loading limit constraint
\eqref{eq:thermal}, so we do not consider it further.
Note that the load at each bus $l_i$, specific costs $c_{i,s}$, generation upper
limits $D_{i,s}$, branch loading limits $F_\ell$ and branch reactances $x_\ell$ are
all exogenous data inputs and not subject to optimization in the considerations here.
In all cases here the network is assumed to be connected and only a single time point is considered. Extensions are discussed in the next section.

Finally active power flows on each branch $f_\ell$ are determined by the
$p_i$ and the auxiliary variables $z_a$ through the constraints
\begin{equation}
  f_\ell \equiv f_\ell(p_i,z_a)
\end{equation}
The different formulations of the network equations presented in Section \ref{sec:lpf}
give rise to different formulations of the linear OPF.
Whether we include the flows $f_\ell$ and additional auxiliary variables $z_a$ as optimization
variables has a significant impact on the computational resources needed to solve
the optimization task. In the following we specify the different formulations of the
linear OPF (LOPF) in detail; their properties are summarized in Table
\ref{tab:formulations}. Note that for a uniquely-defined problem, all the formulations deliver the same optimum.

\subsection{Pure Angle formulation}
In the Pure Angle formulation the optimization
problem (\ref{eq:opt-withoutf}) is solved with the voltage angles as auxiliary variables $\{z_a\} = \{\theta_i \} $
subject to the constraints (\ref{eq:maxdispatch}) and
\begin{align}
  & \big| \sum_i (BK^T)_{\ell i} \theta_i \big|  \le F_\ell  & \forall \, \ell &= 1,\ldots,L, \nonumber \\
  & p_i  = \sum_j \Lambda_{ij} \theta_j   & \forall \, i &= 1,\ldots,N, \nonumber \\
  & \theta_0  = 0.
\end{align}
The first equation ensures no branch overloading (note that it is
sparse, inheriting the sparsity of $K$), the second equation is KCL
and in the final equation the phase angle is fixed at the reference
bus, which removes an unnecessary degree of freedom. Here and in the
following the $p_i$ are used as a short-hand notation according to
equation (\ref{eq:pi_vs_li}).

The Pure Angle formulation is used in the free software tools MATPOWER \cite{MATPOWER}
and PYPOWER \cite{PYPOWER}; it is therefore used as the benchmark implementation
against which we compare all other formulations in
Section \ref{sec:results}.

\subsection{Angle+Flow formulation}

For the Angle+Flow formulation of the LOPF the flows ${f_\ell}$ are introduced as
explicit optimization variables and the voltage angles are retained as
auxiliary variables. Hence we have to solve the optimization problem (\ref{eq:opt-withflow})
with $N$ auxiliary variables, $\{z_a\} = \{ \theta_i\}$ subject to the constraints
(\ref{eq:thermal}) and (\ref{eq:maxdispatch}) and the network equations
\begin{align}
  f_\ell & = \sum_i (BK^T)_{\ell i} \theta_i   & \forall \; \ell &= 1,\ldots,L, \nonumber \\
  p_i & = \sum_\ell K_{i\ell} f_\ell    &\forall \; i &= 1,\ldots,N, \nonumber \\
  \theta_0 & = 0.  & & \label{eq:angle}
\end{align}
The introduction of additional optimization variables $f_\ell$ might
appear to be redundant and unnecessary, but it will be shown to cause
a significant speed-up in some cases. This is because modern solvers
have sophisticated algorithms to `pre-solve' solutions and remove
redundancy that may not be obvious.

This formulation has been used in the literature, for example in
\cite{Sharifnia}.

\subsection{Pure PTDF formulation}

In the Pure PTDF formulation no auxiliary variables
are used such that the optimization problem is given by (\ref{eq:opt-withoutf})
subject to the constraints (\ref{eq:maxdispatch}) and
\begin{align}
   & \big| \sum_i PTDF_{\ell,i} p_i \big| \le F_\ell \qquad \forall \, \ell = 1,\ldots,L. \nn \\
   &  \sum p_i = 0.
\end{align}
This formulation minimizes the number of optimization variables, but
suffers from the fact that the matrix $PTDF$ is dense. This generates
a large number of dense inequalities, which can be slow to process for large problems and may make the feasible
space complicated by introducing lots of interdependencies between the
variables.
This formulation has been used in the literature in, for example,
\cite{Hagspiel}. One advantage of this formulation is that the constraints are independent for each line, so that the constraints can also be limited to subsets
of lines. This is useful when it is known in advance which lines are typically constraining.

\subsection{PTDF+Flow formulation}

The PTDF+Flow formulation does not use any auxiliary variables, but
keeps the flows as explicit optimization variables.
Hence we have to solve the optimization problem (\ref{eq:opt-withflow})
subject to the constraints (\ref{eq:thermal}) and (\ref{eq:maxdispatch})
and the network equations
\begin{align}
  f_\ell &  = \sum_i PTDF_{\ell i} p_i \qquad \forall \, \ell = 1,\ldots,L, \nonumber \\
  \sum_i p_i & = 0 .
\end{align}

This formalism was used in \cite{HINOJOSA2016391,HINOJOSA2016139}.

\subsection{Kirchhoff formulation}

The Kirchhoff formulation is a formulation of the LOPF which only
requires the flow variables $f_\ell$ and introduces no additional
auxiliary variables.  The optimization problem is given by
(\ref{eq:opt-withflow}) subject to the constraints (\ref{eq:thermal})
and (\ref{eq:maxdispatch}) and the network equations
\begin{align}
  \sum_{\ell} K_{i\ell} f_\ell & = p_i       & \forall \, \ell &= 1,\ldots,L, \nonumber \\
  \sum_{\ell} C_{\ell c} x_\ell f_\ell & = 0 & \forall \, c &= 1,\ldots,L-N+1. \label{eq:kvlopt}
\end{align}

This method implements the Kirchhoff circuit laws directly
on the flow variables. It has both a small number of variables and
extremely sparse constraints. As discussed in the introduction, this formulation was used in \cite{Carvalho} and also introduced recently for optimal transmission switching \cite{Kocuk16}.

\subsection{Pure Cycle formulation}

The Cycle formulation of the linear load flow problem introduced
in Section (\ref{sec:flow-cycle}) leads to new formulations of the LOPF.
In the Pure
Cycle formulation we solve the optimization problem
(\ref{eq:opt-withoutf}) by adding  $L-N+1$  auxiliary variables $\{z_a\} = \{h_c\}$
for the cycle flows subject to the constraints (\ref{eq:maxdispatch}) and
\begin{align}
  & \bigg| \sum_{i} T_{\ell i} p_i + \sum_c C_{\ell c} h_c \bigg| \le F_\ell
              \qquad \forall \, \ell = 1,\ldots,L, \nonumber \\
  & \sum_{\ell} C_{\ell c} x_\ell \bigg[ \sum_i T_{\ell i} p_i + \sum_{c'} C_{\ell c'} h_{c'} \bigg]  = 0
    \nonumber \\
      & \hspace{4cm} \forall \, c = 1,\ldots,L-N+1, \nonumber \\
  & \sum_i p_i = 0 .
\end{align}
If $L < 2N$, which is typically true for power networks, this involves
both fewer variables and fewer constraints than the Pure Angle
formulation. However, because for some lines the matrix $T_{\ell i}$
may have many entries, the constraints can only be considered
semi-sparse.

\subsection{Cycle+Flow formulation}

In the Cycle+Flow formulation we add auxiliary variables $\{z_a\} = \{h_c\}$ and include the flow variables $f_\ell$ as explicit optimization variables. The optimization problem is then given by
(\ref{eq:opt-withflow}) subject to the constraints (\ref{eq:thermal}) and (\ref{eq:maxdispatch})
and the network equations
\begin{align}
  & f_{\ell}  = \sum_{i} T_{\ell i} p_i + \sum_c C_{\ell c} h_c  & \forall \, \ell &= 1,\ldots,L, \nonumber \\
  & \sum_{\ell} C_{\ell c} x_\ell f_\ell  = 0 \nonumber   & \forall \, c &= 1,\ldots,L-N+1, \\
  & \sum_i p_i = 0. &&
\end{align}

\section{Extensions to LOPF}\label{sec:extensions}

In this section we briefly sketch some extensions of the
LOPF problem to related problems for which the methodology also applies.

\subsection{Multi-period optimization}

Inter-temporal aspects of optimal power flow, such as the operation of
storage units or power plant unit commitment, can be considered using
multi-period OPF \cite{Wood14,ConvexOptimization}. For periods labeled $t$ with weighting $\pi_t$ the objective function becomes
\begin{equation}
 \min_{\{d_{i,s,t}\}, \{z_{a,t} \}, \{f_{\ell,t}\}} \left[\sum_{i,s,t} \pi_t  c_{i,s} d_{i,s,t} \right] .
\end{equation}
The network flow constraints repeat for each period $t$.

Storage introduces inter-temporal constraints that ensure that the storage state of charge $soc_{i,s,t}$ stays below the maximum energy storage capacity $SOC_{i,s}$:
\begin{align}
  soc_{i,s,t} & = soc_{i,s,t-1} + \eta_{1} d_{i,s,t,\mathit{charge}} - \eta_{2}^{-1} d_{i,s,t,\mathit{discharge}} \nonumber \\
0 \leq  soc_{i,s,t} & \leq SOC_{i,s}   \hspace{1cm} \forall\, i,s,t
\end{align}
The efficiencies $\eta_1, \eta_2$ determine the losses during charging and discharging, respectively.

\subsection{Generation investment optimization}\label{sec:genopt}

For generation investment optimization, the power plant capacities $D_{i,s}$ are promoted from
exogenous parameters to optimization variables with capital costs $C_{i,s}$ \cite{ConvexOptimization}. The objective function becomes
\begin{equation*}
 \min_{\{D_{i,s}\},\{d_{i,s,t}\}, \{z_{a,t} \}, \{f_{\ell,t}\}} \left[\sum_{i,s} C_{i,s} D_{i,s} + \sum_{i,s,t} \pi_t  c_{i,s} d_{i,s,t} \right] \, .
\end{equation*}
The optimization is carried out over multiple periods $t$ representing
different demand and weather conditions, which makes such problems
computationally challenging.

For investment
optimization it is common to approximate the line outage contingency constraints by a
blanket factor, e.g. limiting loading to 70\% of thermal limits, to
reduce the computationally complexity
\cite{DENAII,Hagspiel,Brown2016,Lumbreras201776}.

\subsection{Security-Constrained LOPF}

In Security-Constrained LOPF (SCLOPF) line outages are modelled
explicitly. It is required that no lines become overloaded if there is
an outage of any branches in a critical subset \cite{Wood14}.

SCLOPF can be implemented either by adding to the LOPF problem copies of all the network variables and constraints for networks without the critical branches, or by using Line Outage Distribution Factors (LODFs).

In the LODF formalism, for each branch $k$ which is critical, the following set of constraints are added to the LOPF
\begin{equation}
  |f_\ell^{(k)}| = | f_\ell + LODF_{\ell,k} f_k | \leq F_\ell \hspace{1cm} \forall \ell \neq k \label{eq:lodf}
\end{equation}
Here $f_\ell,f_k$ are the flows before the outage and $f_\ell^{(k)}$
is the flow on $\ell$ after the outage of branch $k$. The flows before
and after the outage are related linearly by the LODF matrix, which
can also be computed efficiently using cycle flows \cite{ronellen16b}.

In the first version of SCLOPF with copies of the network constraints with outages, it is expected that all the benefits of the cycle methods are preserved. In the LODF formalism the density of the LOPF matrix may blunt the benefits of a sparse formulation. The trade-offs between these issues will be examined in a forthcoming paper.

\begin{table*}[!t]
  \caption{LOPF speed-up versus the Pure Angle formulation ($>1$
    means faster), best formulation marked green, worst marked red}
  \label{tab:results}
  \small
	\centering \rowcolors{4}{}{gray!10}
    \begin{tabular}{@{}llrrrrrrr@{}}
      \toprule
      &       & Avg. solution time &                \multicolumn{6}{l}{Speed-up compared to Pure Angle}  \\
&       & (24 periods) &   Angle+   &     Pure      &      PTDF+     &           &   Pure         &  Cycle+          \\
  &         &     Pure Angle [s] &                      Flow & PTDF & Flow & Kirchhoff & Cycle & Flow \\
mode & case &                    &                                 &           &           &           &            &            \\
\midrule
p & case118 &               0.20 &                            1.13 & \slowest     0.24 &      0.53 & \fastest     1.27 &       0.76 &       0.98 \\
  & case300 &               0.45 &                            1.00 & \slowest     0.27 &      0.59 & \fastest     1.12 &       0.60 &       0.67 \\
  & case1354pegase &               1.92 &           \fastest  1.07 & \slowest     0.10 &      0.17 &      0.99 &       0.23 &       0.43 \\
  & case1951rte &               3.21 &                            0.22 & \slowest     0.14 &      0.27 & \fastest     1.30 &       0.32 &       0.55 \\
  & case2383wp &               9.17 &                            0.75 & \slowest     0.27 &      0.44 &  \fastest    1.43 &       0.42 &       0.35 \\
  & case2869pegase &              14.94 &             \fastest               2.19 & \slowest     0.30 &      0.52 &      2.15 &       0.41 &       0.85 \\
  & scigrid &               2.01 &                            1.44 & \slowest     0.10 &      0.19 &  \fastest    1.60 &       0.57 &       1.08 \\
r & case118 &               0.25 &                            0.99 & \slowest     0.12 &      0.23 &  \fastest    1.22 &       0.58 &       0.88 \\
  & case300 &               0.77 &                            1.12 & \slowest     0.11 &      0.20 &  \fastest    1.37 &       0.54 &       0.73 \\
  & case1354pegase &               7.58 &                            1.38 & \slowest     0.06 &      0.10 &  \fastest    2.55 &       0.42 &       0.87 \\
  & case1951rte &              11.96 &                            0.57 & \slowest     0.05 &      0.09 & \fastest     2.70 &       0.46 &       0.93 \\
  & case2383wp &              65.17 &                            3.40 &  \slowest    0.13 &      0.24 & \fastest     4.31 &       1.13 &       1.55 \\
  & case2869pegase &              51.83 &                            0.83 & \slowest     0.06 &      0.10 &  \fastest    3.60 &       0.43 &       1.18 \\
  & scigrid &               3.60 &                            1.62 & \slowest     0.06 &      0.12 & \fastest     2.44 &       0.75 &       1.14 \\
rs & case118 &               0.26 &                            0.99 & \slowest     0.13 &      0.23 & \fastest     1.24 &       0.61 &       0.90 \\
  & case300 &               0.77 &                            1.11 & \slowest    0.11 &      0.19 & \fastest     1.38 &       0.55 &       0.73 \\
  & case1354pegase &               7.45 &                            1.35 & \slowest     0.06 &      0.10 &  \fastest    2.42 &       0.42 &       0.89 \\
  & case1951rte &              11.91 &                            0.58 & \slowest     0.05 &      0.09 & \fastest     2.62 &       0.46 &       0.90 \\
  & case2383wp &              60.73 &                            3.22 & \slowest     0.14 &      0.25 &  \fastest    4.12 &       1.10 &       1.44 \\
  & case2869pegase &              52.88 &                            0.85 & \slowest     0.07 &      0.11 &  \fastest    3.61 &       0.45 &       1.20 \\
  & scigrid &               7.26 &                            2.70 & \slowest     0.12 &      0.25 &  \fastest    4.14 &       1.33 &       2.03 \\
\bottomrule
\end{tabular}
\end{table*}

\section{Results}\label{sec:results}

In this section we compare the computational performance of the
different formulations of the LOPF problem introduced in Section
\ref{sec:lopf} for various different test grids.
All LOPF formulations are implemented in `Python for
Power System Analysis' (PyPSA) \cite{PyPSA}, a free software tool
developed at the Frankfurt Institute for Advanced Studies
(FIAS). The formulation can be changed simply by passing a different argument `formulation' to the LOPF function.
PyPSA is used to generate linear program files (in
CPLEX's .lp format), which are then passed to a linear solver (here we
use the commercial software Gurobi \cite{gurobi}).  The solver is then
run using different algorithms for the linear program (primal and dual
simplex, interior point) and the total solving time averaged over
multiple runs is compared. Only Gurobi's solving time is presented, so that
the results are independent of the program used to generate the optimization problem.
The total solving time includes reading in
the .lp file, pre-solving the matrix system and the solution
algorithm. A computer system with 20  Intel Xeon E5-2650 cores @
2.30GHz each and 128 GB RAM was used for each benchmark.

\subsection{Problem preparation}

Seven different network topologies are considered. case118, case300,
case1354pegase, case1951rte, case2383wp and case2869pegase are taken
from the MATPOWER software package \cite{MATPOWER} test cases (the
IEEE standard cases as well as snapshots from the French TSO RTE and
European networks \cite{RTE-PEGASE}). In addition the open data
SciGRID model of Germany's transmission network \cite{SciGRIDv0.2} is
also tested, which has 585 buses and 948 branches.

Only large networks were considered, because large problems represent
the main target of efforts to improve computational speed. For the
same reason, all networks were tested for multi-period optimization
with 24 hours represented in each problem, which would be typical for
short-term storage optimization or a unit commitment problem. Large
problems also ensure that no small one-off delays can significantly
affect the timing.

Each test grid only has a single snapshot of the load. This was
extended to 24 hours by subtracting a small fraction of normally
distributed random noise $\varepsilon \sim \mathcal N(0, 0.2)$
\begin{equation}
  l_{i,t} = l_{i}\,(1-  |\varepsilon_{i,t}|) ,
\end{equation}
to ensure that the problem remained feasible and the solver was unable
to reduce the problem from 24 identical problems to a single one.

The configuration of the generation was varied in three different
`modes':
\begin{itemize}
\item p: (plain): Only the conventional generators of the model are available. There is no inter-temporal linkage between the snapshots.
\item r: Compared to p, variable renewable generators are added to every single bus to represent
decentralized generation. The time series of the power availability of the renewable generators
are taken at random from wind and time series for Germany for the year 2011 generated using the Aarhus Renewable Energy Atlas \cite{Andresen20151074}.
The renewable generators may be curtailed such that they correspond to dispatchable generators
with no variable costs.
There is no inter-temporal linkage between the snapshots.
\item rs: Compared to r, storage units with a power capacity of a
  third of the nodal mean load are added to the fifteen buses with the
  highest average load. They provide an energy capacity of 6 hours at
  full power capacity and link the snapshots. More than 15 storage units made the computation times intractable.
\end{itemize}

\begin{figure}[t]
  \centering
  \includegraphics[width=\linewidth]{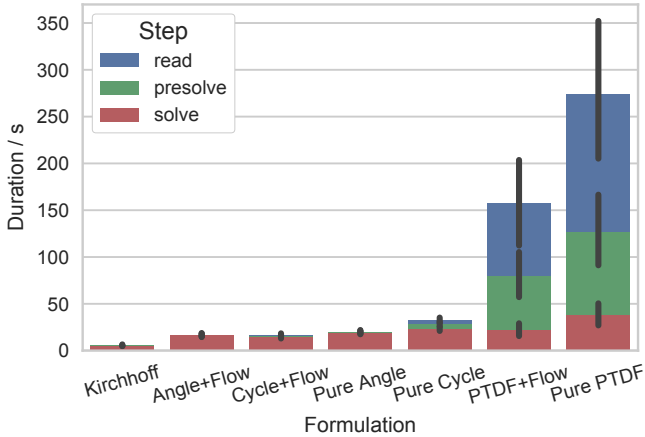}
  \caption{The average duration for each formulation in mode `rs', broken down into the time to read the .lp file, pre-solve the matrix system and solve the problem. The vertical black lines indicate 95\% confidence intervals.}
  \label{fig:duration}
\end{figure}

\begin{figure}[t]
  \centering
  \includegraphics[width=\linewidth]{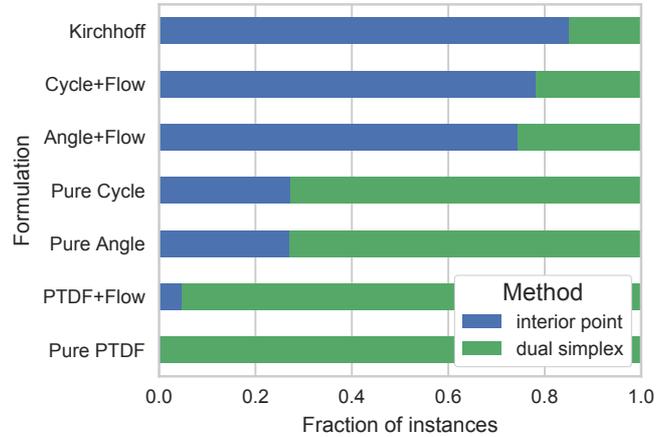}
  \caption{The fastest solution algorithms for each formulation taken over all modes, cases and instances.}
  \label{fig:algorithm}
\end{figure}

For each network, mode and formulation, Gurobi was run in parallel
using the primal simplex, the dual simplex and the interior point
algorithms. The simplex algorithms received one dedicated core each
(they do not work on multiple cores), while the interior point algorithm ran on
two dedicated cores.  The fastest solution was always taken. For each
case and mode combination, 100 instances (i.e. different
randomizations of the load and selections of the renewable time
series) were generated and timed for all formulations except for the
Pure PTDF and the PTDF+Flow formulations. For these only 10 instances
were investigated, since the generation of a single of their lp files
took up to 6 hours. It was checked that all formulations gave
identical results for the same problem, which is to be expected given
that the formulations were shown in Sections \ref{sec:lpf} and
\ref{sec:lopf} to be mathematically equivalent.

The code for running the simulations with
Snakemake \cite{snakemake} is freely available online at
\cite{zenodo}.

\begin{figure*}[t]
  \centering
  \includegraphics[width=\linewidth]{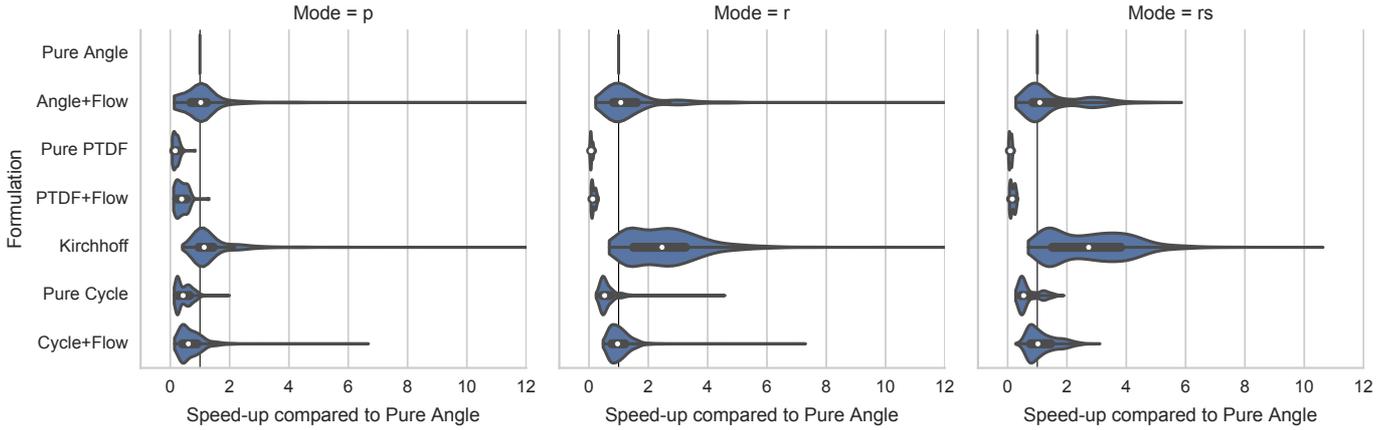}
  \caption{ Speed-up of LOPF compared to Pure Angle for total time (read +
    pre-solve + solve).  The violin plots give the distribution of speed-ups, while the box plots mark the 25\% and 75\% quantiles and the dot marks the median.}
  \label{fig:compare-formulations}
\end{figure*}

\begin{figure}[t]
  \centering
  \includegraphics[scale=0.8]{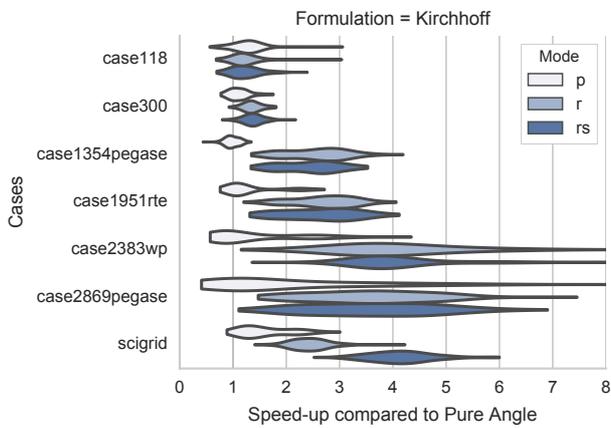}
  \caption{Kirchhoff LOPF speed-up compared to Pure Angle for total time (read +
    pre-solve + solve) per network case.  }
  \label{fig:compare-modes}
\end{figure}

\subsection{Comparing average speed-up of the different formulations}

In Table \ref{tab:results} the speed-up for the different formulations
of the LOPF for the different problems (averaged over 100 instances)
are shown, compared to the standard Pure Angle formulation. The
speed-up is defined by the time taken for the Pure Angle formulation
divided by the time taken for the formulation in question. A speed-up
above 1 means the formulation is faster.

The Kirchhoff formulation is the fastest in all cases where
decentralized renewables are present in the network and the fastest
in all but two cases for the `plain' mode, where the Angle+Flow
formulation is faster by a small margin. For the Kirchhoff formulation
the speed-up factor averages 1.4 in mode `p', 2.6 in mode `r' and 2.8
for mode `rs'. One reason the speed-up is high with renewables is that
the optimization has to weigh up the dispatch at every single bus and
their effects on the flows. A sparser, less interdependent constraint
set is a bigger advantage than in mode `p', where only a few buses
have controllable generators.  Inter-temporal storage introduces even
more interdependences between variables, which again favours the
sparse formulations.

The Angle+Flow formulation is the next fastest, averaging a speed
improvement of 1.11 in mode `p', 1.42 in mode `r' and 1.54 in mode `rs',
despite the fact that there are more variables than the Pure Angle
formulation.

The Cycle+Flow formulation is a factor 0.7 slower than the Pure Angle
formulation in mode `p', but faster by factor 1.04 in mode `r' and
1.16 in mode `rs'. The Pure Cycle formulation is on average slower in
all modes. In particular cases the Pure Cycle formulation is faster
than Pure Angle, but in each of those cases Cycle+Flow is faster.

The PTDF methods are slowest of all, with the Pure PTDF being the
slowest. This is primarily driven by the size of the linear
programming problem file, which takes a long time to read in by the
solver. The size of the file is driven by the dense constraints coming
from the dense PTDF matrix. For large networks with many periods, the
file sizes were many gigabytes, leading to problems writing them and
storing them.  Once the .lp problem is read in and pre-solved, the
solving time is in some cases faster than some of the other methods, a
result also reported by \cite{HINOJOSA2016391,HINOJOSA2016139}.  This
can be seen clearly in the timing breakdown in
Figure \ref{fig:duration}: the average time to solve the problem
for PTDF+Flow is comparable to the Pure Angle formulation once it has
been read in and presolved, but the presolving and reading in add
considerably to the total problem duration. Even comparing just the
`solve' step, Kirchhoff is still faster by a wide margin.

In Figure \ref{fig:algorithm} the fastest solution algorithms for the
linear problems (primal simplex, dual simplex or interior point) are
plotted for each formulation. The Kirchhoff, Cycle+Flow and Angle+Flow
formulations solve in general faster with the interior point
algorithm; the PTDF formulations are faster with the dual simplex. In
no cases was the primal simplex faster.

\subsection{Comparing specific speed-ups of the different formulations}

The average speed-ups of the different formulations in the different
modes masks considerable variations, both between the different
network cases considered and within the instances for each
case. Figure \ref{fig:compare-formulations} shows violin plots of all
the instances and all the cases for each mode and formulation
combination, while in Figure \ref{fig:compare-modes} the different cases
can be seen more clearly for the Kirchhoff formulation.

Consider the speed-up of the Kirchhoff formulation in mode `r' as an
example. The average speed-up is 2.6, but this masks speed-ups for
particular instances that range from a factor 0.7 (i.e. a 30\%
slow-down, for an instance of case118) to factor 20 (for an instance
of case2383wp). Even within a particular case there is significant
variation for particular instances, ranging for case2383wp from 1.2 up
to 20, although with a strong clustering around the mean of 4.3.

\begin{figure}[t]
  \centering
  \includegraphics[scale=0.8]{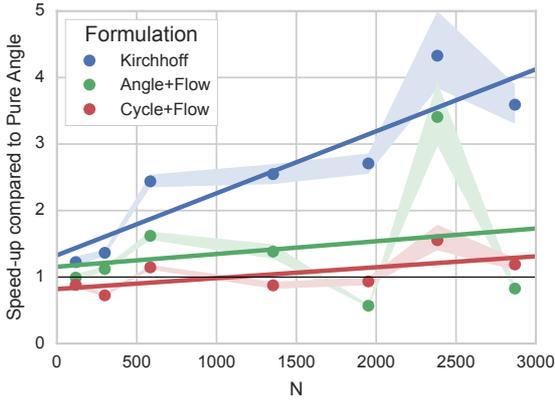}
  \caption{
    Speed-up of LOPF compared to Pure Angle per buses, shown are the
    mean values with 99\% confidence interval and the result of a linear
    regression of all values for the three fastest formulations in mode `r'.
    % refer to http://seaborn.pydata.org/generated/seaborn.tsplot.html#seaborn.tsplot for details
  }
  \label{fig:speedup-N}
\end{figure}

In 3\% of the instances the Kirchhoff formulation in mode `r' is in
fact slower than the Pure Angle formulation, and all these instances
are for the cases with a smaller number of buses, case118 and case300.
Figure \ref{fig:speedup-N}  reveals that this
is part of a bigger trend: In the `r' mode, the Kirchhoff
formulation speed-up grows with the size of the network, measured
in terms of the number of buses. The increase in speed-up with network size also holds true for the `rs' mode.

%It might have been expected that the speed-up scales with the number
%of closed cycles in the network, given that the number of KVL
%constraints in the Kirchhoff formulation \eqref{eq:kvlopt} is the
%number of cycles. In Figure \ref{fig:speedup-per-ratio-ci} the
%speed-up is plotted against the ratio of the number of cycles
%($L-N+1$) to the number of buses. No clear relationship can be
%recognized.

%Similarly there is no relationship between the speed-up and the
%numbers of variables, constraints and non-zeroes in the constraint
%matrices, before and after pre-solving, for any of the cases, modes or
%formulations.

%\begin{figure}[t]
%  \centering
%  \includegraphics[width=7cm]{speedup-L-r}
%  \caption{ Speed-up to pure angles in Kirchhoff `r' versus number of lines }
%  \label{fig:speedup-L}
%\end{figure}

Of all the cases, instances and modes, the Kirchhoff formulation was
fastest in 79.3\% of the problems, while the Angle+Flow was fastest
in 12.5\%, Angle in 7.5\% and Cycle+Flow in 0.7\%. If we
restrict to the modes `r' and `rs', then the Kirchhoff is fastest in
91.6\% of the problems, Angle+Flow in 5.9\%, Angle in 2.1\% and Cycle+Flow in 0.4\%.

\begin{table*}[!t]
  \caption{Speed-up of LOPF with capacity optimization compared to the Pure Angle formulation, best formulation marked green}
  \label{tab:results-capacity}
  \small
	\centering \rowcolors{4}{}{gray!10}
        \begin{tabular}{@{}lrrrrr@{}}
            \toprule
                & Mean solution  & \multicolumn{4}{l}{Speed-up compared to Pure Angle} \\
                                    &  time: Pure                  &                      Angle+           &           &    Pure        &    Cycle+        \\
                formulation &     Angle [s] &                      Flow & Kirchhoff & Cycle & Flow \\
                \midrule
                case118        &               1.00 &                            0.89 & \fastest     1.15 &       0.74 &       0.84 \\
                case300        &              13.60 &                            5.25 & \fastest     5.70 &       2.37 &       3.53 \\
                case1354pegase &             539.93 &                            3.01 & \fastest    12.97 &       2.05 &       4.76 \\
                case1951rte    &             914.55 &                            3.10 & \fastest     9.39 &       1.73 &       4.36 \\
                case2383wp     &            7815.68 &                            1.61 & \fastest    21.47 &       5.91 &      20.59 \\
                case2869pegase &            5172.15 &                            2.23 & \fastest    24.94 &       1.36 &       7.28 \\
                scigrid        &             347.72 &                            2.73 &     10.04 &       3.97 & \fastest     10.26 \\
                \bottomrule
                \end{tabular}
\end{table*}

The high level of variation of the speed-up for different cases and
instances (reflecting different load and renewable profiles) means
that in practice it may be advisable, given a particular problem, to
run several formulations in parallel on a machine with multiple cores and take the solution from
whichever solves first, much as linear program solvers like Gurobi can
be configured to run multiple solution algorithms in parallel, given
the difficulty in predicting the runtime in advance.

\subsection{Generation investment optimization}

\begin{figure}[t]
  \centering
  \includegraphics[scale=0.8]{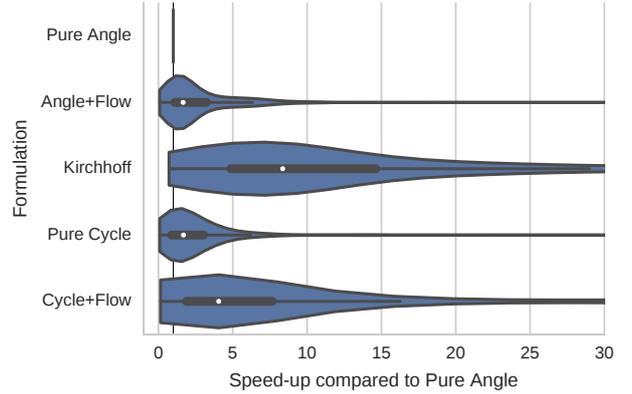}
  \caption{ Speed-up of LOPF with capacity optimization compared to Pure Angle for total time (read +
    pre-solve + solve).}
  \label{fig:compare-formulations-capacity}
\end{figure}

\begin{figure}[t]
  \centering
  \includegraphics[scale=0.8]{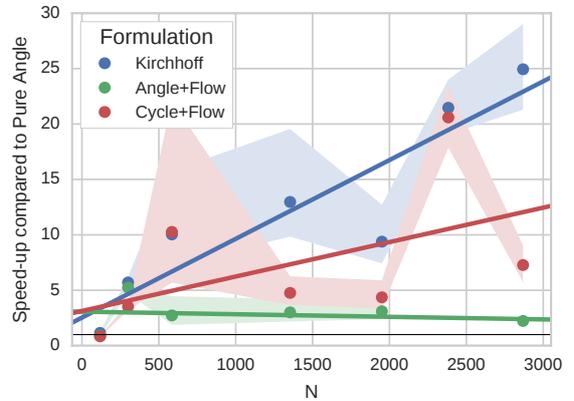}
  \caption{
    Speed-up of LOPF with capacity optimization compared to Pure Angle per buses. Shown are the
    mean values with 99\% confidence interval and the result of a linear
    regression of all values.
    % refer to http://seaborn.pydata.org/generated/seaborn.tsplot.html#seaborn.tsplot for details
  }
  \label{fig:speedup-N-capacity}
\end{figure}

In a final set of computations, the capacities of all generators and
storage units were included in the optimization following Section
\ref{sec:genopt} for the case `rs' with renewables and storage,
optimized over 24 time periods.  With capacity optimization, the
problems take much longer to solve and a time-out of $10^4$ seconds
(just under 3 hours) was set on all calculations, since some instances
were failing to converge in a reasonable time. In the Pure Angle
formulation this limit was hit for some of the larger cases, breaching
the limit in 55\% of the instances for case2383wp, 18\% for
case2869pegase, and 1\% for each of case1354pegase, case1951rte and
scigrid. The Angle+Flow formulation breached the limit in 25\% of
instances for case2383wp. For the cycle-based formulations all
instances solved within the time limit.  The PTDF method was excluded
from this comparison given its slowness in previous results.

The results for the seven test cases are presented in Table
\ref{tab:results-capacity} and graphed in Figure
\ref{fig:compare-formulations-capacity}. Overall the speed-up factors
are higher than for the LOPF without capacity optimization. Once again
the Kirchhoff method is the fastest in most cases, averaging 12 times
faster than Pure Angle over all cases, rising to 25 times faster for
the biggest case case2869pegase. An individual instance of
case1354pegase solved 213 times faster. The Cycle+Flow formulation
performed better than it did for the LOPF without capacity
optimization, solving on average 7.4 times faster than Pure Angle and
faster than Angle+Flow in most cases. The Cycle+Flow formulation was
on average the fastest for the scigrid network, with an individual
instance of the scigrid network finishing 388 times faster than Pure
Angle.

Once again there is a trend for the speed-up to be higher with the
Kirchhoff method the more nodes there are in the network, see Figure
\ref{fig:speedup-N-capacity}. This will benefit exactly the cases
which take a long time to solve.

In these calculations only 24 time periods were included for the
optimization. In general more periods are necessary to account for
different weather conditions, which pushes computation times from
hours to days. It is expected that the Kirchhoff method will thus make
possible calculations that were not even possible with the Pure Angle
formulation.  For example, in \cite{Hoersch2017} some of the authors
considered the joint optimisation of generation, storage and
transmission capacities for networks with 362 nodes over 2920
representative time periods. With the Kirchhoff formulation, the
problems solved within an average of 10.2 hours; with the Angle
formulation none of the optimisations converged within four days, at
which point the calculations were broken off.

%\subsection{Comparing performance for different networks}

%Take a vertical slice across Table \ref{tab:results} for the Kirchhoff formulation; compare to number of cycles, etc.

%The speed-up is fairly non-deterministic (i.e. does not seem to follow a regular pattern, e.g. correspondence to number of cycles).

%This is similar to LP problems in general - it is hard to predict in
%advance which problems solve faster and which method works better on
%individual problems, apart from trying them out.

\section{Conclusion}\label{sec:conclusions}

In this paper a new formulation of the linear optimal power flow
(LOPF) problem, the Cycle formulation, has been presented. The new
formulation uses a graph-theoretic decomposition of the network into a
spanning tree and closed cycles; this results in both fewer decision
variables and fewer constraints in the LOPF problem.

A comprehensive study of the numerical performance of different LOPF
formulations has been provided by applying them to computationally
challenging problems such as multi-period LOPF with storage dispatch
and generation capacity expansion. While for many problems the new
Cycle formulation was faster than the traditional LOPF formulation in
terms of voltage angle variables, it was in most (but not all) cases
out-performed by another cycle-based formulation, the Kirchhoff
formulation.  The Kirchhoff formulation implements the two Kirchhoff
circuit laws directly on the flow variables, resulting in performance
which is considerably faster than the standard Angle formulation used
in today's power system tools. Both cycle-based formulations show the
greatest speed-up in very large networks with decentralized
generation, which are exactly the kinds of problems that are becoming
increasingly important with the rise of distributed renewable
energy. In the Kirchhoff formulation the LOPF can solve up to 20 times
faster for particular cases, while averaging a speed-up of approx.~3
for the networks considered in this paper.  In 92\% of the problems
with distributed generation, the Kirchhoff formulation was the fastest
formulation.  If generation capacities are also optimized, the average
speed-up rises to a factor of 12, reaching up to factor 213 in a
particular instance. In a small number of specific cases the Cycle
formulation was the fastest.

Future further applications of cycle-based formulations could include
the transmission expansion problem, stochastic optimization and the
application of graph decomposition to the full non-linear optimal
power flow problem.

\section*{Acknowledgments}
We gratefully acknowledge support from
the German Federal Ministry of Education and
Research (BMBF grant nos.~03SF0472A-E) and
the Helmholtz Association (joint initiative `Energy System 2050 -- a contribution of the research field energy' and
grant no.~VH-NG-1025 to D.W.).
The work of H. R. was supported in part by the IMPRS Physics of Biological
and Complex Systems, G\"ottingen.

% --- Literatur -------------------------------------------------------------------

\bibliographystyle{elsarticle-num}
\biboptions{sort}
\bibliography{lopf}

\end{document}